\pdfoutput=1
\documentclass{nature}

\usepackage{csquotes} 

\usepackage{hyperref}
\usepackage{amssymb, amsmath, mathptmx} 

\usepackage{graphicx}
\usepackage[font=small,labelfont=bf]{caption}
\usepackage[version=4]{mhchem} 
\usepackage[usenames, dvipsnames]{color} 

\makeatletter
\newcommand*{\rom}[1]{\expandafter\@slowromancap\romannumeral #1@}
\makeatother

\title{Superferromagnetism and domain-wall topologies in artificial \enquote*{pinwheel} spin ice}

\author{Yue Li$^1$, Gary W. Paterson$^{1, *}$, Gavin M. Macauley$^1$, Fabio S. Nascimento$^2$, Ciaran Ferguson$^1$, Sophie A. Morley$^{3,4}$, Mark C. Rosamond$^5$,  Donald A. MacLaren$^1$, Rair Mac\^edo$^1$, Christopher H. Marrows$^3$,  Stephen McVitie$^{1, *}$ \& Robert L. Stamps$^{1, 6, *}$}

\begin{document}

\maketitle

\begin{affiliations}
 \item SUPA, School of Physics and Astronomy, University of Glasgow, Glasgow, G12 8QQ, United Kingdom.   
 \item Departamento de F\'isica, Universidade Federal de Vi\c cosa, Vi\c cosa, 36570-900, Minas Gerais, Brazil.
 \item School of Physics and Astronomy, University of Leeds, Leeds, LS2 9JT, United Kingdom.
 \item Department of Physics, University of California, Santa Cruz, California, 95064, USA.
 \item School of Electronic and Electrical Engineering, University of Leeds, Leeds LS2 9JT,  United Kingdom.
 \item Department of Physics and Astronomy, University of Manitoba, Manitoba, R3T 2N2, Canada.
\end{affiliations}

\begin{abstract}
For over ten years, arrays of interacting single-domain nanomagnets, referred to as artificial spin ices, have been engineered with the aim to study frustration in model spin systems.  Here, we use Fresnel imaging to study the reversal process in \enquote*{pinwheel} artificial spin ice, a modified square ASI structure obtained by rotating each island by some angle about its midpoint. Our results demonstrate that a simple 45$^\circ$ rotation changes the magnetic ordering from antiferromagnetic to ferromagnetic, creating a superferromagnet which exhibits mesoscopic domain growth mediated by domain wall nucleation and coherent domain propagation. We observe several domain-wall configurations, most of which are direct analogues to those seen in continuous ferromagnetic films.  However, novel charged walls also appear due to the geometric constraints of the system. Changing the orientation of the external magnetic field allows control of the nature of the spin reversal with the emergence of either 1-D or 2-D avalanches. This unique property of pinwheel ASI could be employed to tune devices based on magnetotransport phenomena such as Hall circuits.

\end{abstract}

Artificial spin ice (ASI) systems have been used not only as a route to new physical phenomena, but also to gain insight into fundamental physics. 
Such capabilities are only possible because these structures are able to emulate the behaviour of assemblies of the individual spins in atomic systems. This is done by controlling the shape and size of each nanoelement to ensure that they behave as single-domain magnets. 
One of the most appealing and perhaps well known aspects of ASI systems is their capability to display geometrical frustration. 
This magnetic topological frustration gives rise to interesting properties\cite{Wang2006, Perrin2016, Wang2016, Zeissler2013, Mengotti2011, Drisko2015, Zhang2013, Drisko2015, Gilbert2015, Gilbert2014, Lammert2010, Silva2013, Ladak2011}, such as monopole-like defects\cite{Mengotti2011, Silva2013, Ladak2011, Branford2012, hugli2012artificial, gliga2013spectral} and multifold ground-state degeneracy\cite{Wang2006, Perrin2016, shi2017frustration, gilbert2014emergent}.

The classic artificial spin ice tiling, that of square ice, has a well known long-range antiferromagnetic ground state arising from its two-fold degenerate \enquote*{two-in-two-out} spin configuration of each vertex\cite{morgan2011thermal, Zhang2013, Farhan2013}. 
This structure, which obeys the so-called \enquote*{ice rule}\cite{Pauling1935} and possesses four well-defined vertex energies, was initially investigated by Wang et al~\cite{Wang2006}. 
Their work ignited great interest in not only square ASI, but also in several other ASI arrangements. 
In particular, Morrison et al.~\cite{morrison13} pointed out the importance of vertex of interactions and their dependence on geometry. 
A simply modified square ASI system provides a recent example of emergent dynamics: the \enquote*{pinwheel} ice \cite{gliga2017emergent, Rair2017pinwheel}. 
The pinwheel geometry is obtained by rotating each island in square ASI around its centre\cite{Rair2017pinwheel}. 
Gliga et. al. have found that thermal relaxation in this system behaves as if it obeyed an intrinsic chirality \cite{gliga2017emergent}. 
Frustration in pinwheel ASI is markedly different than that in square ice in that the energies of the different pinwheel units are found to be nearly degenerate, whereas the energy levels of square ice vertices are well separated\cite{Rair2017pinwheel}.

In this work, we use Lorentz transmission electron microscopy (LTEM)\cite{mcvitie15} to directly visualise the magnetisation reversal process in a pinwheel ASI array in the presence of a static externally applied magnetic field.
Under such conditions, the system behaves as a superferromagnet, i.e. an ensemble of macrospins with collective ferromagnetic behaviour~\cite{bendanta07}. Our superferromagnets has coherent domain growth and shrinking as opposed to the chain avalanche reversal seen in square ASI~\cite{morgan11a,morgan13}. 
The different magnetic domains seen in pinwheel ASI are separated by domain walls, some of which behave much like the classical ferromagnetic N\'eel domain walls in continuous films. 
However, new types of charged domain walls are also observed and the magnetic charge ordering of these walls is dependent on the magnetisation alignment of the neighbouring domains. 
The behaviour of these walls and domains is significantly affected by the field orientation, which is also investigated.
These properties of pinwheel ASI offer a possible avenue to designing functional materials exploiting the emergent magnetic spin textures and controllable reversal dimensionality.

\section*{\large Results}
\begin{figure}[h!]
	\centering
	\includegraphics[width=8cm]{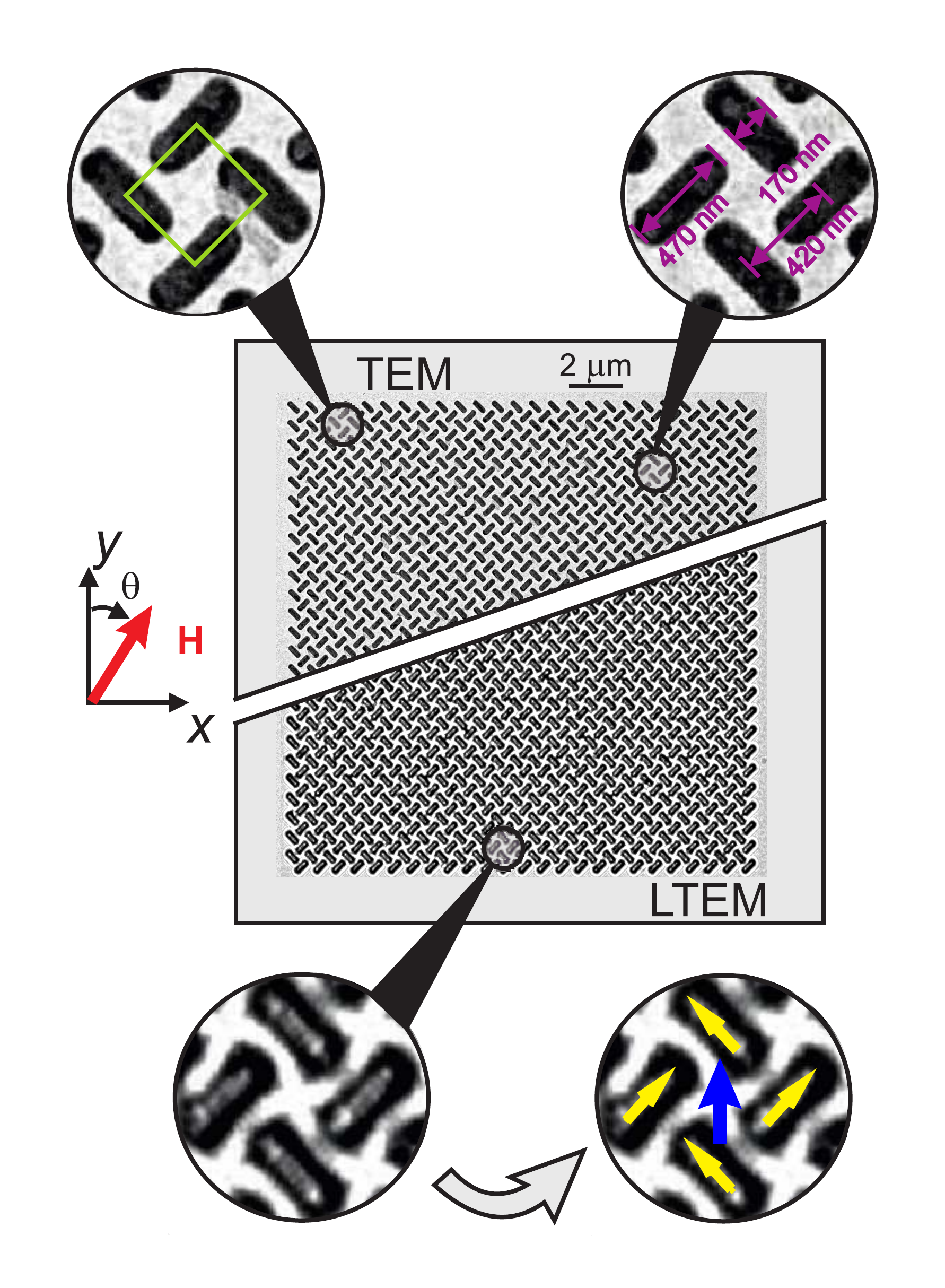}
	\caption{\textbf{Example of in-focus TEM and LTEM images of an artificial pinwheel spin ice array.} The top-left part shows an in-focus TEM image of a pinwheel ASI array composed of two interleaved 25$\times$25 sublattices. To the left we show the field direction defined with respect to the array edges. The top-left inset is the zoomed-in image of a diamond pinwheel unit and the top-right inset shows the in-plane dimension of each island and the centre-to-centre distance between nearest-neighbour islands. The bottom-right part of the array is an LTEM image of the same pinwheel array with an enlarged view of a single unit (the bottom-left inset). The small yellow arrows in the bottom-right inset illustrate the magnetic moment directions identified from the image contrast, and the large blue arrow shows the net magnetisation direction of the unit.}
	\label{PASI_TEM} 
\end{figure}

Array edges for pinwheel ASI arrays are typically either \enquote*{diamond} or \enquote*{lucky-knot} designs\cite{Rair2017pinwheel}, where the array termination edges lie at either 45$^\circ$ or 90$^\circ$ to an island long axis, respectively. 
Here, we investigate the behaviour of a diamond-edge permalloy (\ce{Ni_{80}Fe_{20}}) pinwheel array of two interleaved collinear $25\times25$ sublattices as shown in the in-focus TEM image in the upper half of Fig.~\ref{PASI_TEM}.
For this edge type, the element centres define a diamond geometry\cite{Rair2017pinwheel} (as shown in Fig.~\ref{PASI_TEM}'s top left inset). 
Each individual nanomagnetic island is 10~nm thick, 470~nm long, and 170~nm wide, with a centre-to-centre separation between nearest-neighbour islands of 420~nm, as shown the top right inset to Fig.~\ref{PASI_TEM}.

\section*{Ising Hysteresis Behaviour}

\begin{figure}[h!]
\centering
\includegraphics[width=0.9\linewidth]{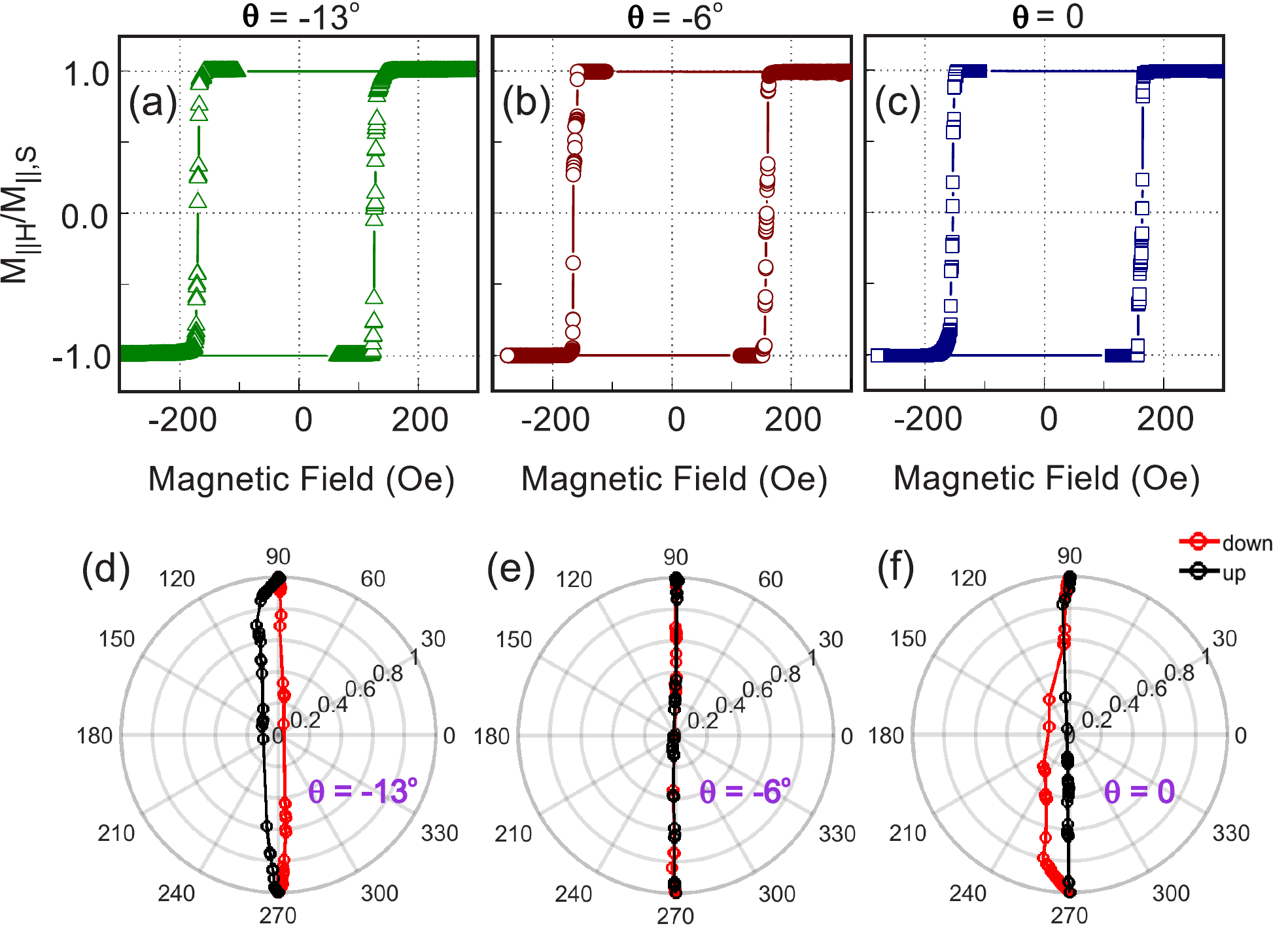}
\caption{\textbf{Array net magnetisation reversal of the artificial pinwheel spin ice array}. (a)-(c) Normalised component of the magnetisation aligned with the field, \textit{$M_\parallel$}, during a field sweep for \textbf{(c)} $\theta = -13^\circ$, \textbf{(d)} $\theta = -6^\circ$ and \textbf{(e)} $\theta = 0$. \textbf{(d)-(f)} Normalised net moment of the entire array displayed in polar coordinates, with the field aligned at the same angles as those in \textbf{(a)-(c)}. The \enquote*{down} (H \textless 0) and \enquote*{up} branches are indicated by the symbol colour (red and black) and arrow directions.}
\label{Mag_orient}
\end{figure}

In order to characterise the behaviour of the pinwheel ASI, we first look at the behaviour of the Ising net magnetisation of the individual pinwheel units, where each unit is formed by the four nearest-neighbour islands.
This is done by examining the defocused Fresnel LTEM images recorded during a reversal, an example of which is shown in the bottom half of Fig.~\ref{PASI_TEM}. 
Magnetic contrast arises through deflection of the electron beam by the induction from the magnetisation of each island. 
Since the magnetisation lies along the island's long axis, a bright edge will be seen on one side and a dark edge on the other, with the direction dependent on the orientation of the moment. 
From this, the magnetisation of each unit can be directly measured through its magnetic contrast as shown in the inset of Fig.~\ref{PASI_TEM}.
In this way, it is then possible to follow the moment orientations throughout the entire array as a function of the external field. 
Example Ising hysteresis loops extracted from these orientations are given in Figs.~\ref{Mag_orient}(a)-(c) for various field angles with respect to the array edges, as defined in Fig.~\ref{PASI_TEM}. 
Although the coercivity for each field angle is slightly different, the general behaviour is similar. 
Note that when the external static field lies parallel to the $y-$axis ($\theta = 0$), the component of the externally applied field along the easy-axis of each of the pinwheel islands is the same.
The measured coercive field, $H_C$, agrees very well with that calculated using the field protocols outlined in supplemental (see Fig.~\ref{S1}). 
However, despite extensive efforts to replicate the precise details of the magnetisation, this model does not adequately capture the richness of the behaviour discussed in the following sections. 

For an array of entirely uncoupled islands, the easy anisotropy axis should lie at $\theta = 0$. 
Inter-island interaction may modify this angle in a real system, so it is important to characterise it experimentally.
We can determine the anisotropy axis of our array by examining the net magnetisation component perpendicular to the field. 
This component is small at low angles of applied field, and in fact shows an interesting dependence on the applied field angle.
This can easily be seen if the $x-$ and $y-$components of the net magnetisation along the array axes are plotted in polar coordinates. 
In Figs.~\ref{Mag_orient}(d)-(f), we re-plot in this way the data of Fig.~\ref{Mag_orient}(a)-(c) ($-13^\circ$, $-6^\circ$ and $0$).
In order to understand the meaning of these plots, is useful to think of the pinwheel array as a two interpenetating sublattices of collinear islands~\cite{gliga2017emergent} as shown in Fig.~\ref{Edge_effect1}(b).
With the external field applied parallel to the easy anisotropy axis of the array, the reversal of each sublattice happens simultaneously, causing the polar hysteresis loop to collapse, and the reversal should be described by overlapping lines to and from 90$^\circ$ and 270$^\circ$. 
For the pinwheel array, this occurs at $\theta = -6^\circ$ as shown in Fig.~\ref{Mag_orient}(b).
When the field is misaligned with the easy axis, the magnetisation rotates with a sense of direction that reflects the sign of the angle, $\theta$.
For example, in Fig.~\ref{Mag_orient}(a) for $\theta = -13^\circ$, the moment rotates clockwise, whereas in Fig.~\ref{Mag_orient}(c) for $\theta = 0$, the moment rotates anticlockwise.

Using the width of the polar hysteresis loop as a measure of the angle between the applied field and anisotropy axes, we estimate that the anisotropy axis lies at $-5.7^\circ \pm1.4^\circ$ to the array edge (see Supplementary Fig.~\ref{Edge_effect} for further details).
Careful measurement of in-focus TEM images of an untilted array confirms that the angles between the sublattices and with respect to the array edges in the realised array are accurate to within $\pm 0.6^{\circ}$.
While this interesting result deserves further investigation, it is beyond the scope of this work and, other than the angle offset, we do not expect it to dramatically affect the reversal process which will be discussed in the following sections.

\section*{Magnetisation Reversal and Domain Formation}

The hysteresis loops shown in Figs.~\ref{Mag_orient}(a)-(c), constructed using the component of magnetisation parallel to the field, $M_{\parallel}$, suggests a ferromagnetic ordering of the array. 
Due to the mesocopic scale inherent to ASI, we can examine the Ising magnetisation of individual islands as a function of the externally applied field.
As is commonly done in ASI, for the reminder of this work we will consider the array as a system of four-island units as shown in the insets of Fig.~\ref{PASI_TEM}. 
Here and elsewhere we adopt the unit type names from square ASI vertices\cite{budrikis12}, as detailed in Fig.~\ref{vertex_infor}. 
So far we have only discussed the ferromagnetic behaviour associated with small angles of the externally applied static field.
We find that the magnetic ordering and reversal behaviour at high angles of applied field are markedly different.

\textit{\textbf{Reversal at Low Applied Field Angles}}

\begin{figure}[h!]
\centering
\includegraphics[width=\linewidth]{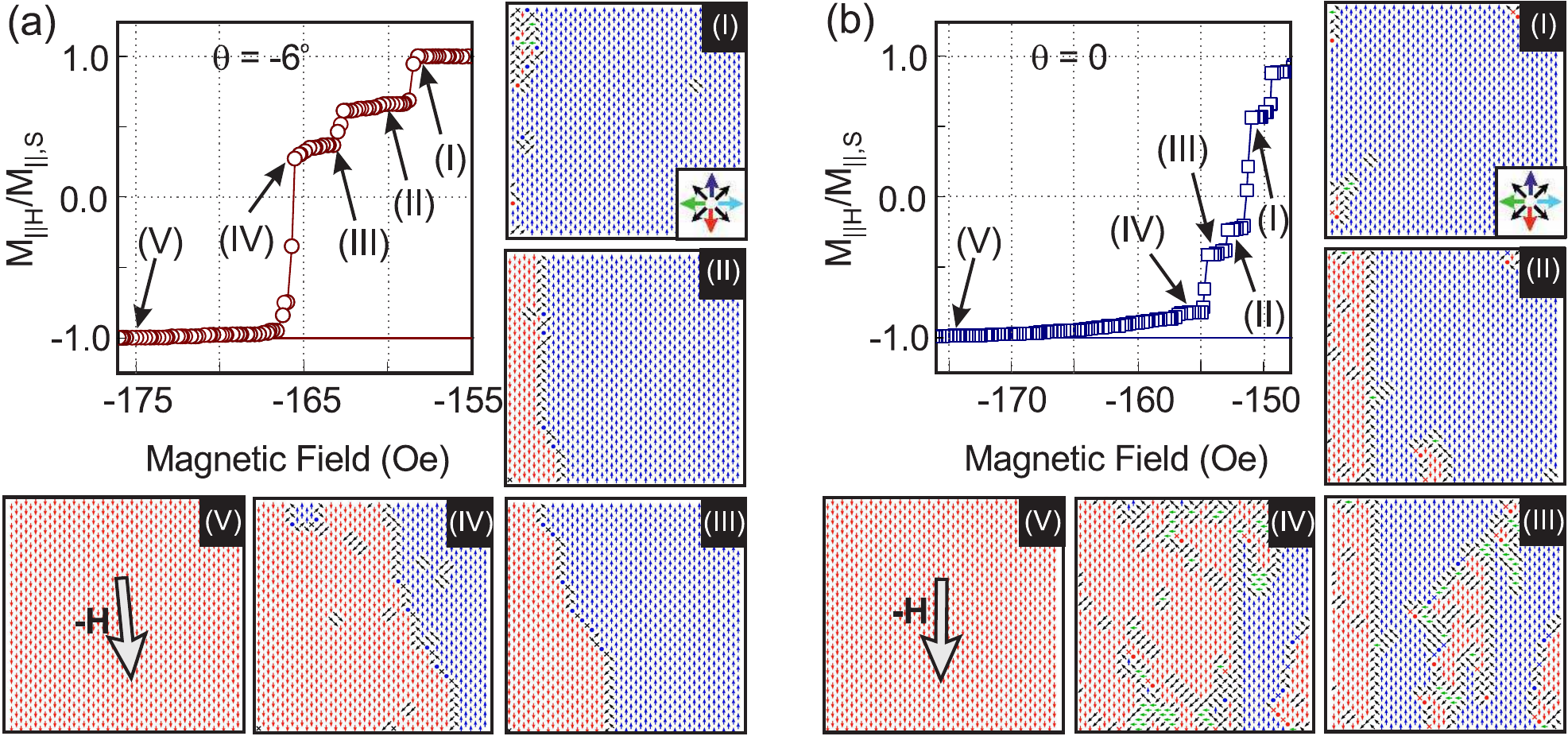}
\caption{\textbf{Mapping the field-driven evolution of domains}. Field-induced domain growth and domain wall patterns in an entire ASI array with applied field angles of \textbf{(a)} $\theta = -6^\circ$ and \textbf{(b)} $\theta = 0$. For both cases, five points ((I)-(V)) are marked in the hysteresis loops across the reversal. We also give snapshots of the net moment of the pinwheel units composing the array at each point marked. The unit magnetisation orientation is depicted by colour-coded arrows, as shown by the arrow colour wheel inset in (I). Further information on the net moment and magnetic charge for all possible domain and domain wall unit configurations is provided in Fig.~\ref{vertex_infor}. }
\label{DomainPattern}
\end{figure}

In Fig.~\ref{DomainPattern}, we show examples of field-driven evolution of the four island unit magnetisation configuration composing an entire pinwheel array in the vicinity of coercive field.
The colours represent the unit magnetisation direction as defined in the legend. 
We show snapshots at different field angles in (a) and (b) at various field magnitudes (points marked in the hysteresis loops going from (I) to (V)).
Further details can be seen in the supplemental videos of the full reversal and snapshots of individual island reversal (see supplemental Video S1 and supplemental Fig.~\ref{MagIslandPattern}). 
Here, we only focus on the general domain reversal behaviour which can be observed from the colour contrast.

When the array is saturated (e.g. as marked as (V) in Fig.~\ref{DomainPattern}), a single mesoscopic domain is formed by the so-called Type \rom{2} units. 
The Type \rom{2} unit possesses the largest net moment, as one might expect, and zero net magnetic charge.
At the small angles of applied field shown in Fig.~\ref{DomainPattern}, reversal starts through a small number of nucleation points, typically located at the edge of the array where the element reversal energy is lower, and progresses by domain growth through domain wall movement perpendicular to the direction of the field.
We note that the behaviour of magnetisation reversal at low field angles mimics that observed for easy-axis reversal of continuous ferromagnetic films  with uniaxial anisotropy\cite{Yulan2015}. 
Interestingly, the reversal appears somewhat more ordered at $\theta = -6^\circ$ than at $\theta = 0$.
This angle offset is consistent with the analysis results of the previous section which showed that the easy anisotropy axis for this array lies at around -6$^\circ$.

\textit{\textbf{Reversal at High Applied Field Angles}}

At higher angles of applied field with respect to the array edge, the reversal process is quite different. 
This is because, as $\theta$ is increased from 0 to 45$^\circ$, the easy axis of one sublattice and the hard axis of the other one will become more closely aligned with the field.
Therefore, one sublattice will switch before the other one during a reversal. This gives rise to a `ratcheting' behaviour yielding a stepped hysteresis loop as shown in Fig.~\ref{DomainPattern30}(a) for an applied field angle $\theta = 30^\circ$. 
Further details can be seen in the supplemental videos of the full reversal and snapshots of individual island reversal (see supplemental Video S2). 
We note that there is some discrepancy between coercive fields within the positive and negative range at this field angle.
The coercive field varies between all field sweeps and we attribute this to variation in the precise magnetisation configuration during reversal and to possible small sample movements during the measurement changing the applied field angle with respect to the array.

\begin{figure}
\centering
\includegraphics[width=14cm]{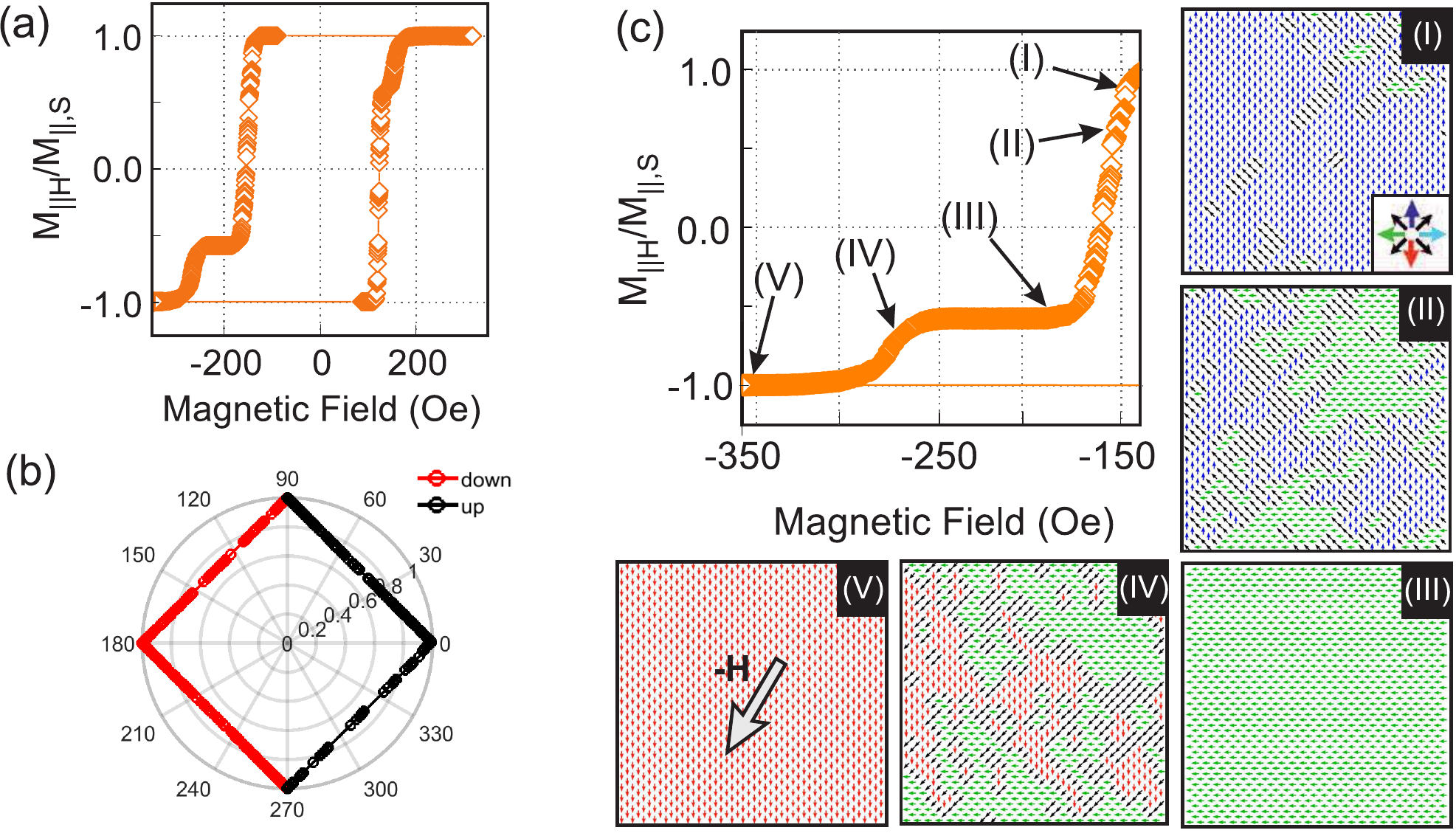}
\caption{\textbf{Mapping the field-driven evolution of domains at high field angles}. Hysteresis behaviour of the \textbf{(a)} net magnetisation component parallel to the externally applied field direction and \textbf{(b)} the net magnetisation with respect to the array edges in polar form, with the field applied at an angle of 30$^\circ$. \textbf{(c)} Field-induced domain growth and domain wall patterns in an entire ASI array. The unit magnetisation orientation is depicted by the colour-coded arrows as defined in the inset to part \rom{1} of \textbf{(c)}.}
\label{DomainPattern30}
\end{figure}

At sufficiently high angles of applied field, the large difference in the applied field angle with respect to the easy axes of the two sublattices causes one sublattice to \emph{completely} reverse before the other one starts.
Because the shape anisotropy of an island only allows the moment to align parallel or antiparallel to the long-axis, the array net magnetisation is constrained to move along 45$^\circ$ lines.
This can be seen for the data in Fig. \ref{DomainPattern30}(a) in the polar plot of the same data in Fig \ref{DomainPattern30}(b). 
Starting with the moment pointing up, when the first sublattice reverses, the moment moves from north to west, and when the second sublattice reverses, the net moment changes from west to south, and so on. 
Thus, the hysteresis loops at high angles of applied field describe a rotated square with a sense of direction that reflects the misalignment angle between the field and the anisotropy axis.

The behaviour of the hysteresis loops shown at higher angles of applied field can be translated into a reversal process mediated through a different mechanism to that at low field angle. 
This can be seen in the texture of the magnetisation across the reversal shown in the snapshots of the magnetisation in Fig.~\ref{DomainPattern30}(c). 
In this case, islands in the sublattice with their easy axes more closely aligned with the field are more likely to reverse first, forming diagonal stripe patterns.
Examples of this can be seen in panels (\rom{1}) and (\rom{2}) of Fig.~\ref{DomainPattern30}(c). 
As the nanomagnets do not couple strongly to those in adjacent diagonal lines, reversal of the entire array occurs through many nucleation points, creating a spatially inhomogeneous reversal with scattered stripe domains.
When one sublattice completely switches (e.g. at point (III) in the hysteresis plot in Fig.~\ref{DomainPattern30}(c)), the fully magnetised net magnetisation lies perpendicular to the initial domain direction.
The process then repeats for the other sublattice, to complete the reversal.
As a consequence of different reversal mechanisms, the critical field angle marking the transition between the square and \enquote*{stepped} loops, can be determined from the relative populations of domain walls. 
This is described in the next section.

\section*{Mesoscopic domain-wall topologies}
\begin{figure}
\centering
\includegraphics[width=\linewidth]{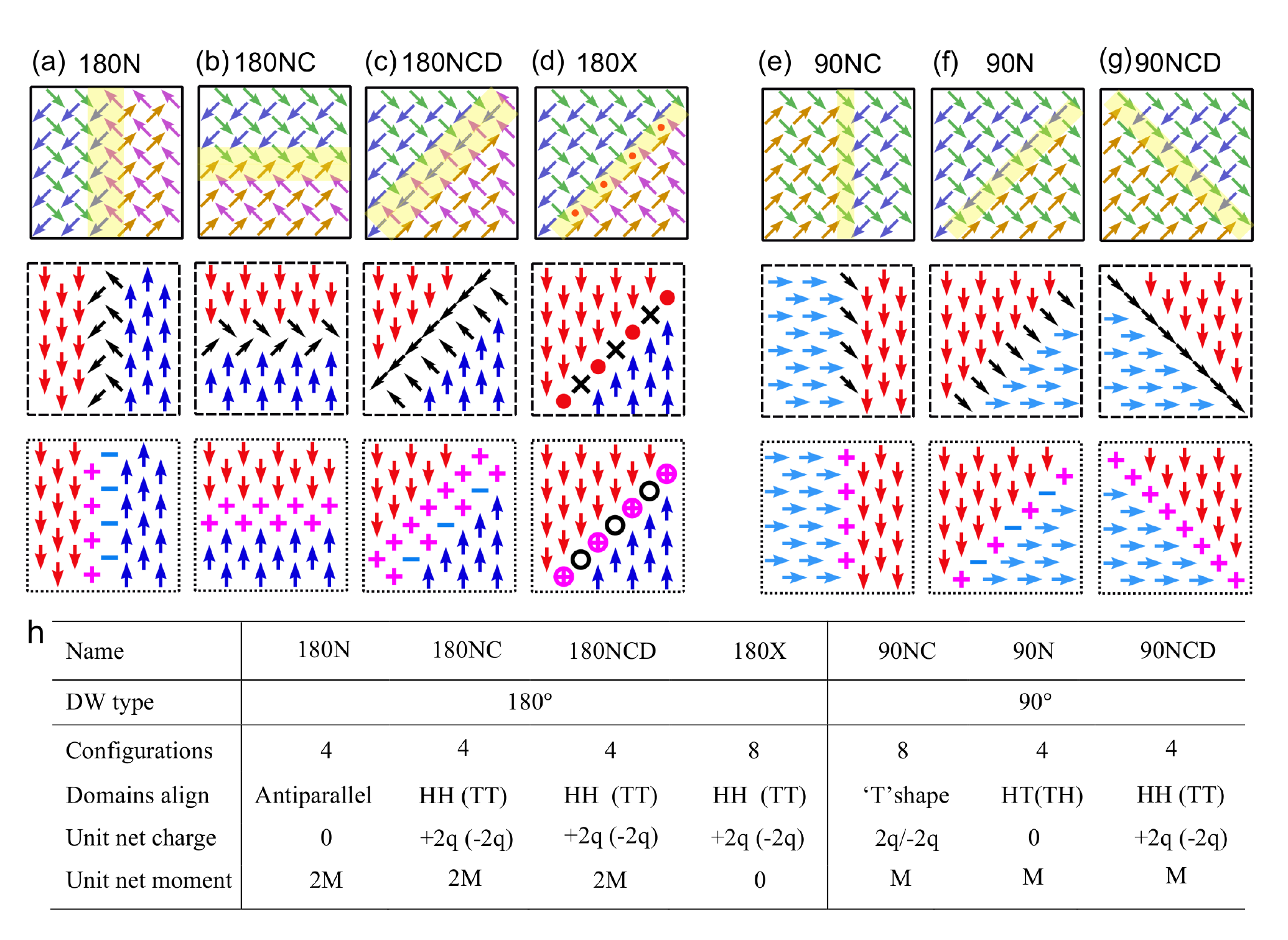}
\caption{\textbf{Mesoscopic domain wall magnetisation and charge-ordering topologies}. \textbf{(a-d)}, Schematic examples of the possible $180^\circ$ DW configurations in pinwheel ASI containing four categories of domain wall (DW): \enquote*{180N} \textbf{(a)}, \enquote*{180NC} \textbf{(b)},  \enquote*{180NCD} \textbf{(c)} and  \enquote*{180X} \textbf{(d)}. \textbf{(e-g)} Schematics of possible $90^\circ$ DW configurations consisting of three types: \enquote*{90NC} \textbf{(e)}, \enquote*{90N} \textbf{(f)} and  \enquote*{90NCD} \textbf{(g)}. The smaller arrows of the configuration at the top, framed by the black solid box, represent the magnetic moments of islands with domain walls highlighted in yellow. The larger arrows in the middle and bottom row images indicate the unit moments of the domains and walls. The bottom images (with dotted frames) show the net charge distributions of the same DWs, where the \enquote*{+/-} signs reveal the net magnetic charges of Type \rom{3} units with $\pm$2q, the \enquote*{$\oplus/\ominus$}symbols indicate the charged Type \rom{4} units with $\pm$4q and the open black circles represent the zero-charge Type \rom{1} units. The net charge of each unit is determined by the dipole magnetic charges of the island using dumbbell model (see Figure~\ref{vertex_infor}\textbf{(d)}). \textbf{(h)}, Summary of DW features in which \enquote*{HH}, \enquote*{TT}, \enquote*{HT} and \enquote*{TH} are short for \enquote*{head-to-head}, \enquote*{tail-to-tail}, \enquote*{head-to-tail} and \enquote*{tail-to-head}, respectively.}. 
\label{DWs}
\end{figure}

In the reversal processes that were described in the previous section, large domains are seen at low angles of applied field, separated by transition regions, much like domains and domain walls in continuous ferromagnetic films. 
In pinwheel ASI, the domain walls separate neighbouring mesoscopic domains and each wall type exhibits a discrete macrospin texture.
In the reversal regime seen at low applied field angles, the domains are almost entirely formed by Type \rom{2} units grouped together throughout a reversal.
These carry the largest moments which appear in the macroscale as ferromagnetism.
Within a domain wall, the macrospin texture is composed by the arrangements of either Type \rom{3} or a mixture of Type \rom{4} and \rom{1} units.
These can be identified by seven classes of domain walls which are depicted in the columns of Figure~\ref{DWs}(a)-(g), where the top row shows one possibility for a magnetic island configuration, the middle row its equivalent unit magnetisation, and the bottom row the unit charge determined using the dumbbell model\cite{Castelnovo2007} as explained in Figure~\ref{vertex_infor}(d). 
The \enquote*{+/-} symbols represent Type \rom{3} units with two positive / negative net magnetic charges; \enquote*{$\oplus$} / \enquote*{$\ominus$} indicates the positive / negative Type \rom{4} units possessing four charges; and \enquote*{$\circ$} are the uncharged Type \rom{1} units.

Each class of domain wall has multiple element configurations.
For example, in Figure~\ref{DWs}(a), two neighbouring domains possess two possible alignments, where the net magnetisations are antiparallel, and the domain wall has two possible magnetisation directions (pointing to the left or right).
As a consequence, this domain wall has four possible magnetisation configurations.
The number of all possible wall configurations is indicated in the third row of the table in Figure~\ref{DWs}(h) (other possible wall configurations are given in Figure~\ref{DW2}).
The seven domain walls can be categorised by the angle between magnetisation orientations of the adjacent domains into 180$^\circ$ or 90$^\circ$ domain walls.
All walls can be further categorised by the alignment of the adjacent domains: either antiparallel, head-to-head (HH), tail-to-tail (TT), or head-to-tail (HT) and vice versa; and by the net charge and moment of a unit, as indicated by the remaining rows of the table.

Drawing from continuous film ferromagnets, we designate all walls in pinwheel ASI by the angle between adjacent domains followed by the minimum number of letters denoting the wall type. 
For the 180$^\circ$ walls (Figures~\ref{DWs}(a)-(d)), the walls are N\'eel (\enquote*{N}), charged N\'eel (\enquote*{NC}), diagonal charged N\'eel (\enquote*{NCD}) and diagonal cross-tie walls (\enquote*{X}).
The 180N walls are analogues of a classical N\'eel wall\cite{Middelhoek1963, Lee2009}, which is uncharged, and the 180X wall resemble a cross-tie wall\cite{Lohndorf1996} formed by alternating Type \rom{4} and Type \rom{1} units. 
The 180NC and 180NCD walls are charged walls not seen continuous ferromagnetic films which lack the reduced degrees of freedom of our pinwheel lattice. 
We note, however, that analogous domain wall configurations are commonly observed in highly anisotropic continuous structures such as nanowires\cite{petit2010magnetic, o2008direct}.

The polarity of charged walls depends on the magnetisation orientation of the adjacent domains.
For example, the domain wall carries positive or negative net unit charge when the magnetisation directions of neighbouring domains are head-to-head or tail-to-tail, respectively.
This is analogous to the characteristic signatures of charged walls in ferroelectric materials\cite{meier2015functional, bednyakov2015formation} in which walls carry polarised electrostatic charges.

All 90$^\circ$ walls (Figure~\ref{DWs}(e)-(g)) separate domains in which the magnetisation directions lie at right angles to one another, and all exhibit N\'eel rotation.
Following the 180$^\circ$ wall naming system, we denote these as \enquote*{90NC}, \enquote*{90N} and \enquote*{90NCD}.
The 90N wall, being uncharged, is analogous to a classical N\'eel wall.
As in the 180$^\circ$ walls, charged N\'eel walls exist which are not found in natural ferromagnets due to the energetically unfavourable head-to-head (see Figure~\ref{DWs}) or tail-to-tail (see Figure~\ref{DW2}) alignment.
The charge ordering of this wall type is also found to be dependent on the magnetisation orientation of their adjacent domains.
These peculiar properties of specific charge ordering in pinwheel ASI are the direct result of the high anisotropy within a system of discrete magnetisation.

\begin{figure}
\centering
\includegraphics[width=0.5\linewidth]{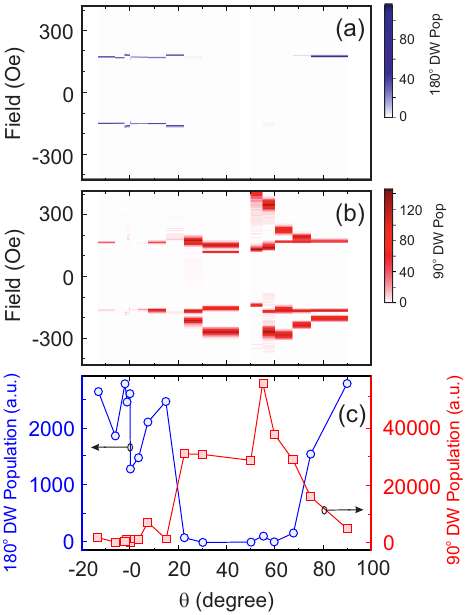}
\caption{\textbf{Population statistics for $180^\circ$ and $90^\circ$ domain walls}. \textbf{(a)} $180^\circ$ and \textbf{(b)} $90^\circ$ DW total populations as a function of the magnetic field strength in a full hysteresis loop as a function of the applied field angle, $\theta$. \textbf{(c)} Field-angle-dependence of the accumulative $180^\circ$ and $90^\circ$ DW populations across the field sweep of the hysteresis loops shown in \textbf{(a)}. The domain wall motifs are defined in Fig.~\ref{DWs_unit}.}
\label{DW_Pop}
\end{figure}

At the level of individual islands, the fundamental difference between the formation of 180$^\circ$ and 90$^\circ$ walls arises from individual moment reversals of the two sublattices, as illustrated by the top row images of Figure~\ref{DWs}(a)-(g).
When one goes through the domain wall interface of a 180$^\circ$ wall, the spins in both sublattices reverse simultaneously, whereas for a 90$^\circ$ wall, only spins in one of the sublattices flip.
This fact can be used to map the transition between the reversal regimes of pinwheel ASI, from ferromagnetic ordering at low angles of applied field to the spatially inhomogeneous reversal at higher angles of applied field.
As the applied field angle increases from zero, one sublattice easy axis becomes more aligned with the field, while that of the other lies at a higher angle to the field and, thus, the two sublattices are no long coupled and a transition from 180$^\circ$ walls to 90$^\circ$ ones should occur.

To characterise the extent of the ordered and the spatially inhomogeneous reversal regimes, we examine the domain wall population statistics across a M-H loop as a function of the applied field angle by counting the number of times a domain wall \enquote*{motif} appears at each applied field strength. 
This result is shown in Figs.~\ref{DW_Pop}(a) and (b) for 180$^\circ$ and 90$^\circ$ walls, respectively. 
The domain wall motifs represent the smallest cluster of spins that define each wall (see sketch in the supplemental Fig.~\ref{DWs_unit}), and these are composed by multiple pinwheel units such as those discussed so far in the context of unit net moment (see insets in Fig.~\ref{PASI_TEM}). 
From Figs.~\ref{DW_Pop}(a)-(b) we can infer that the reversal process at low angles happens mostly though the formation of 180$^\circ$ walls and it extends from at least -13$^\circ$ (the minimum angle measured) to $\sim 20^\circ$.
The 180$^\circ$ walls mostly disappear at higher applied field angles (from $\sim$ 20$^\circ$ to $\sim$ 70$^\circ$), giving way to the formation of 90$^\circ$ walls (examples of all individual wall motif populations are given in the supplementary Fig.~\ref{DW_Pop_absolute}).
Reversal in the latter, spatially inhomogeneous regime clearly happens in a two-step process where one sublattice reverses before the other, as shown in Fig.~\ref{DomainPattern30}. 
The minimum absolute coercive field does not change dramatically across all angles and regimes because the shape anisotropy of the individual island is the main determinant of this property and of the anisotropic nature of the array, with inter-island coupling having a smaller but important effect.
To visualise the regime angular dependence more easily it is useful to build a domain wall population count for the full M-H loop as a function of angle. 
In Fig.~\ref{DW_Pop}(c) we show the sum of all 180$^\circ$ and all 90$^\circ$ wall units for each full loop at each applied field angle from -13$^\circ$ to 90$^\circ$ and, indeed, we see a preferential behaviour of these domain walls for the same range of angles seen in Figs.~\ref{DW_Pop}(a)-(b) before the integration across the M-H loop.

\section*{\large Discussion and Conclusions}
Pinwheel artificial spin ice provides an example of how a simple geometry modification can dramatically affect the magnetic properties of a spin ice array. 
Here, we have shown experimentally the emergence of superferromagnetism in this structure.

We expect array edges to influence aspects of the magnetisation such as array anisotropy and that the extended nature of the island may also have some effect. For instance, recent theoretical work on the pinwheel geometry has shown that the details of the thermal ground state depend on the array edges \cite{Rair2017pinwheel}. 
In order to probe this experimentally, preliminary measurements comparing magnetisation processes in diamond pinwheel arrays with different edge `cuts' have been made. 
The results are suggestive that there may be an effective magnetic anisotropy axis dependence on array edge geometry (See Supplemental Fig.~\ref{Edge_effect1}). 
Definitive conclusions on this aspect require a more detailed and extensive experimental study that goes beyond the scope of the present work.

The reversal process is strongly affected by the direction of the field with respect to the array edges. 
In pinwheel ASI, the dimensionality of magnetic avalanches in the reversal process is determined by the the field angle $\theta$. 
In the low-field-angle regime ($\lesssim$ 20$^\circ$ from array edges), the magnetic ordering is ferromagnetic.
Reversal of the mesoscopic domains in this regime is through two dimensional avalanches of macrospins (see supplemental Fig.~\ref{MagIslandPattern}), while at applied field angles approaching 45$^\circ$ the magnetisation is disordered and reversal is through the formation of 1-D stripes. 
The low applied field angle behaviour is the opposite of the magnetisation processes in square ASI, where 1-D monopole-like defects and dirac-like strings form~\cite{Mengotti2011, Silva2013, hugli2012artificial, gliga2013spectral, pollard2012propagation, gilbert2015direct}, and is a direct result of modification of the inter-island coupling.
In square ASI, the strongest coupling is with the nearest neighbours and the coupling strength falls off monotonically with increasing distance~\cite{morley2018effect}. 
By rotating each island in square ASI by 45$^\circ$, the nearest-neighbour coupling that is dominant in square ASI is greatly reduced and the dipolar coupling strength increases with distance, peaking at the fourth nearest neighbour~\cite{Rair2017pinwheel}. 

The pinwheel ASI system is prone to the formation of domain walls analogous to those seen in continuous film natural ferromagnetic materials, such as N\'eel and cross-tie walls.
However, novel and intriguing domain walls types can also be seen. 
These have specific charge ordering and net moments due to the high anisotropy and constrained degrees of freedom of the system. 
Furthermore, we have shown that by simply changing the orientation of the externally applied field with respect to the array edges it is possible to completely modify the nature of the domain wall configurations and the field evolution. 
This unique property of pinwheel ASI could be used to effectively tune devices based on magnetotransport phenomena such as the recently suggested Hall circuits~\cite{chern17}.

Lastly, the work reported here has concentrated on one particular array geometry. 
The key driving force behind the interest in other ASI systems has been the tunability of key magnetic properties through the geometrical design. 
We expect this also to be true for the superferromagnetic pinwheel ASI, where the coupling parameters can be varied so that different phases can be achieved or controlled.
Our results show that this structure presents an interesting model system for experimental exploration of fundamental magnetisation processes such as magnetic interfaces, exchange bias phenomena, and spin wave propagation.
For example, the array geometry may be tailored in such a way as to extend the Ising-like domain walls reported here to spread over several elements, potentially leading to controlling over the mesoscopic wall formation and propagation.

\section*{\large Methods}

\subsection{\textbf{Sample Fabrication}}
The permalloy (\ce{Ni_{80}Fe_{20}}) ASI arrays were patterned on electron-transparent silicon nitride ($\ce{Si_3N_4}$) membranes using electron-beam lithography and lift-off metallisation. The Si$_{3}$N$_{4}$ membranes were spin coated with ZEP520A: anisole (1:1) with film thickness $\approx$ 140~nm samples at 4 krpm for 40 s and baked at 180 °C for 180 s. The spin ice arrays were then defined using electron beam lithography using an electron dose of 343~$\mu$C/cm$^{2}$. The pattern was developed for 70~s in N50 solution. Ni$_{80}$Fe$_{20}$ was evaporated with thickness 10 nm into the pattern and lifted off in micro-posit remover 1165 at 70 °C.

\subsection{\textbf{LTEM Measurement}}
All experimental results in this work are from Fresnel imaging of ASI arrays in the magnetic field produce by the partially excited objective lens of a JEOL ARM200CF transmission electron microscope.
In Fresnel imaging, the Lorentz force deflection of an electron-beam by the integrated induction produced by each island creates bright and dark edges along the long axis in a defocused image from which the direction of magnetisation can be inferred\cite{Williams2009, Qi2008}.
The defocus in the experiments was 5~mm.
An example of a Fresnel image of saturated pinwheel ASI is shown in the lower left part of Figure~\ref{PASI_TEM}, where the arrows in the bottom inset indicate the magnetisation of each of four islands composing a pinwheel unit.
In the experiments, the in-plane component of the magnetic field was varied by tilting the sample between $\pm 25^\circ$ and $\mp 25^\circ$ in the 700~Oe magnetic field of the objective lens, $H_{obj}$, while a 10~fps video was recorded to track the evolution of the magnetisation.
The videos, and example of which is included in supplemental, were then processed to extract the net magnetisation direction of each island, assuming a single-domain exists and the island acts as a macrospin.

\pagebreak
\bibliography{ref}
\bibliographystyle{naturemag}

\section*{\large Acknowledgements}
This work was supported by the Engineering and Physical Sciences Research Council (EPSRC Grant Nos. EP/L002922/1 and EP/L00285X/1) and the University of Glasgow.
Y. Li was funded by the China Scholarship Council.
G. W. Paterson, S. McVitie and D. A. MacLaren received partial support from EPSRC Grant No. EP/M024423/1. 
G. M. Macauley received support from the Carnegie Trust for the Universities of Scotland. 
R. Mac\^edo was supported by the Leverhulme trust.
R. L. Stamps acknowledges the support of the Natural Sciences and Engineering Research Council of Canada (NSERC) -
R. L. Stamps a \'et\'e financ\'ee par le Conseil de recherches en sciences naturelles et en g\'enie du Canada (CRSNG).
F.S.N. has benefited from discussions with A. R. Pereira.

\section*{\large Author Contributions}
Y. L., G. W. P. and S. McV. designed and performed the experiments and analysis, and interpreted the results. G. M. M., F. S. N. and R. M. worked on the theoretical aspects. C. F., S. A. M. and M. C. R. fabricated the samples under the supervision of C. H. M. and D. A. M.. Y. L., G. W. P. and R. M. wrote the manuscript, with contributions from all authors. F. S. N. and R. L. S. conceived the geometry. R. L. S. conceived this work. All authors contributed to interpreting the results.


\clearpage
\section*{\LARGE Supplemental Material}

\setcounter{figure}{0}
\setcounter{table}{0}
\setcounter{section}{0}
\setcounter{page}{1}
\renewcommand{\thefigure}{S\arabic{figure}}
\renewcommand{\thetable}{S\arabic{table}}
\renewcommand{\thesection}{S.\arabic{section}.}

\renewcommand{\theHtable}{Supplement.\thetable}
\renewcommand{\theHfigure}{Supplement.\thefigure}
\renewcommand{\theHfigure}{Supplement.\thesection}


\section*{Calculations} 

\begin{figure}[ht]
\centering
\includegraphics[width=8cm]{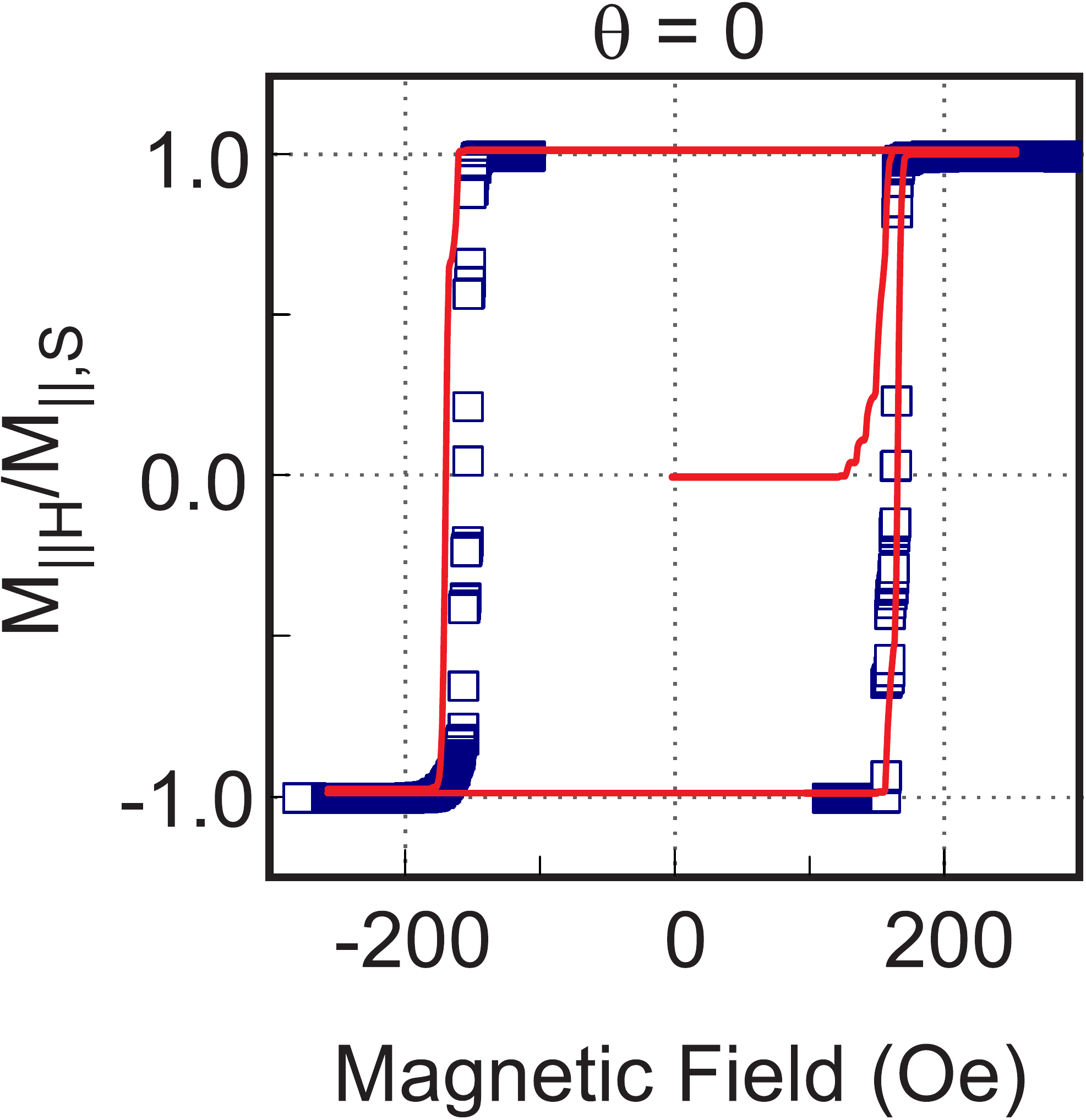}
\caption{Experimental (blue squares) and calculated (red line) normalised component of the magnetisation aligned with the field, \textit{$M_\parallel$}, during a field sweep for $\theta = 0$. }
\label{S1}
\end{figure}

A comparison between calculated and experimental Ising M-H loops is shown in Fig.~\ref{S1}.
The average experimental coercive field value is $|H_C| = 158\pm 8$~Oe, which compares very well with the simulation value of 164~Oe.

We approximate each extended island as an Ising point dipole with an energy barrier, $H_C$, taken from micromagnetic simulations . In hysteresis calculations, our switching criterion is given by \cite{porro2017magnetization ,budrikis12,budrikis12a}
\begin{equation}
-({H}_{dip}^{(i)}+H_{ext})\cdot \hat{\sigma}_i>{H}_{C}^{(i)}.
\label{hz}
\end{equation} 
where $H_{dip}$ is the total dipolar field from all other spins acting on spin $i$\cite{Rair2017pinwheel} and $\vec{H_{ext}}$ is the externally applied magnetic field  which is allowed to rotate with respect to sample by an angle, $\theta$, as defined in Fig.\ref{PASI_TEM}. This means that the component of the total field lying antiparallel to an island's magnetisation must be greater than the island's intrinsic switching field, ${H}_{C}^{i}$, in order for that island to flip.

For the energy barrier calculation we employ the software MuMax3. We use the same island dimensions as those in our experimental array. 
The cell size in each direction was taken as 2.5~nm; the saturation magnetisation was taken to be 860~$\times$ 10$^{3}$ A/m; and the exchange stiffness was 13.0~pJ/m with a damping 0.002.

\section*{Anisotropy Axis Orientation} \label{AnisotropyAxis}
\begin{figure}
\centering
\includegraphics[width=14cm]{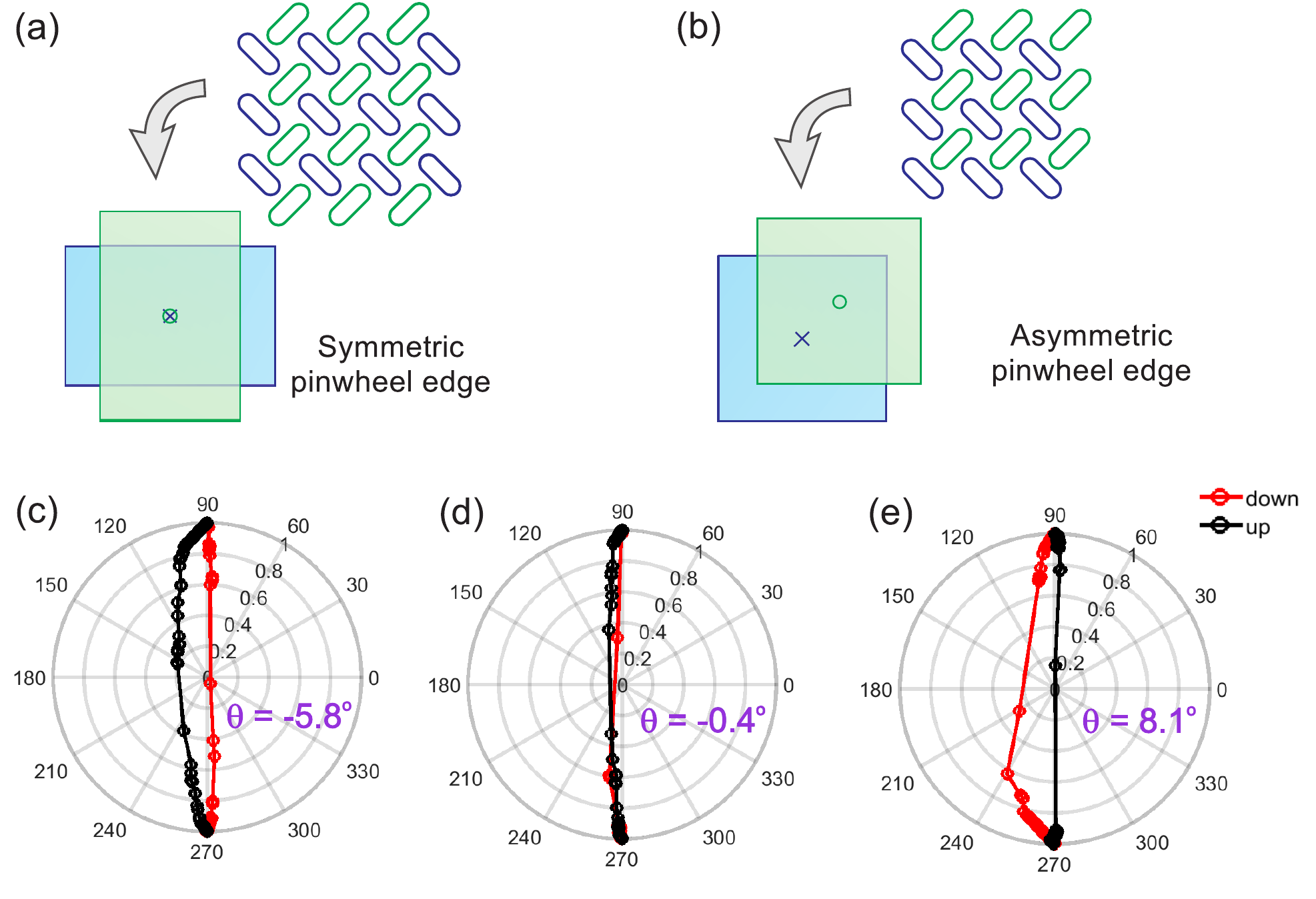}
\caption{\textbf{Edge types and anisotropy axis orientation:} \textbf{(a)} Symmetric and \textbf{(b)} Asymmetric diamond edge pinwheel tilings. \textbf{(c)-(e)} Net Ising moment orientation of the entire array with the symmetric boundary in polar coordinates, with the field aligned at \textbf{(c)} $\theta = -5.8^\circ$, \textbf{(d)} $\theta = -0.4^\circ$, and \textbf{(e)} $\theta = 8.1^\circ$ with respect to the array edge. }
\label{Edge_effect1}
\end{figure}

\begin{figure}
\centering
\includegraphics[width=14cm]{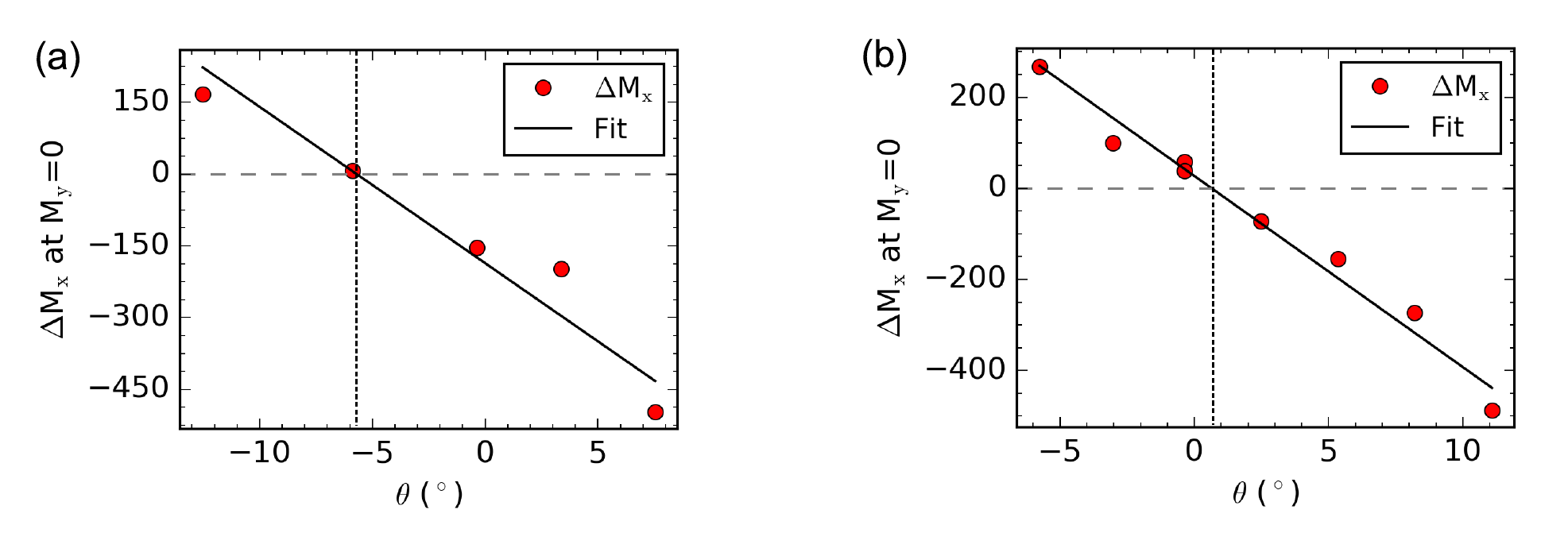}
\caption{\textbf{Anisotropy axis determination:} The separation between up and down loops at $M_y$ = 0, $\Delta M_x$, as a function of the field angle for the \textbf{(a)} asymmetric- and \textbf{(b)} symmetric-edged pinwheel arrays. Fits (solid lines) to the data (symbols) reveals that the field angle at which the $\Delta M_x$ is zero, corresponding to the magnetic anisotropy axis of the arrays, are $-5.7^\circ \pm 1.4^\circ$ for the asymmetric boundary array and $0.7^\circ \pm 0.4^\circ$ for the array with the symmetric boundary.}
\label{Edge_effect}
\end{figure}

It is important to note that whilst we have worked with a diamond-edge array throughout this work, it is possible to `cut' the pinwheel structure in two different ways: the one discussed in the manuscript, where the centre of both sublattices do not line up (see Fig.~\ref{Edge_effect1}(b)); and another one where the centres of both sublattices lie in the same position (see Fig.~\ref{Edge_effect1}(a)). 
We call these `asymmetric' and `symmetric' pinwheel, respectively. 
Whilst these edges are not expected to significantly impact the hysteresis behaviour of large arrays, they may play some role in determining the anisotropy axis of the array.
The results of initial experimental measurements of a symmetric array are shown in Figs.~\ref{Edge_effect1}(c)-(e). 
In the symmetric array, the anisotropy axis, as indicated by the angle at which the sublattices reverse simultaneously resulting in a closed polar hysteresis loop, is found to occur at $\theta = -0.4^\circ$.
The precise angle of the anisotropy axis, determined through examining the angular dependence of the polar hysteresis loop width as discussed in the main text, is 0.7$\pm$0.4$^\circ$ as shown in Fig.~\ref{Edge_effect}(b).
Further work is required to fully understand this observation.

\newpage
\section*{Square ASI Vertices to Pinwheel Units Labeling} \label{label}

\begin{figure}
	\centering
	\includegraphics[width=\linewidth]{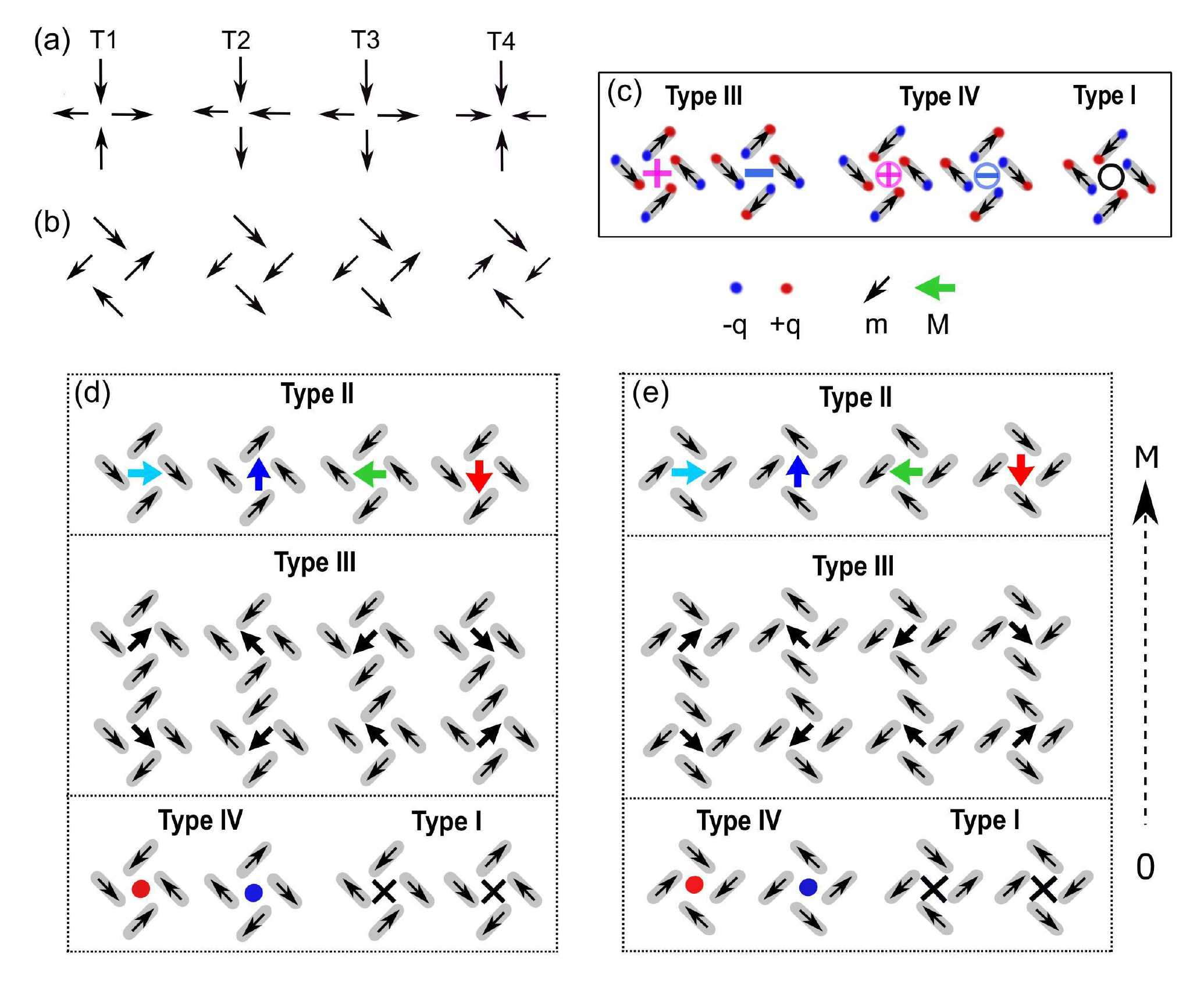}
	\caption{\textbf{Vertex origin and categories in pinwheel ASI}. Examples of magnetisation topologies for T1, T2, T3 and T4 square ASI vertices \textbf{(a)} turned into  pinwheel ASI units \textbf{(b)}. 
 \textbf{(c)} Net charge of each units, determined by the dipole charges of the magnetic element using the dumbbell model.
 \textbf{(d)} Sixteen possible pinwheel unit configurations grouped by their net moment and  \textbf{(e)} the matching set of configurations for units of opposite chirality.}
	\label{vertex_infor}
\end{figure}

The definition of all pinwheel units is shown in Fig.~\ref{vertex_infor}. 
The unit type names in the pinwheel geometry are carried over from the square ASI vertex definition as shown in panels (a) and (b). 
In (c) we show the net charge of each unit as determined from the dumbbell model. 
Finally, Fig.~\ref{vertex_infor}(d) and (e) show the net moments of all possible pinwheel unit configurations.

For Type \rom{3} units, the symbols \enquote*{+} and \enquote*{-} represent +2q and -2q, \enquote*{$\oplus$} and \enquote*{$\ominus$} represent the +4q and -4q of Type \rom{4} units.
Open black circles represent Type \rom{1} vertices with zero charge.
Uncharged Type \rom{2} units have the largest magnetisation (M).
Type \rom{3} units have a charge of $\pm$2q and a smaller net magnetic moment of $\sqrt{2}/2$M.
The fully charged ($\pm4$q) Type \rom{4} units are indicated by filled circles, where red and blue are the positive and negative charge, respectively.
Uncharged Type \rom{1} units are indicated by black crosses, and have zero moment.

\newpage

\section*{Domain and Domain Wall Formation} \label{DWform}

\begin{figure}[ht]
	\centering
	\includegraphics[width=0.9\linewidth]{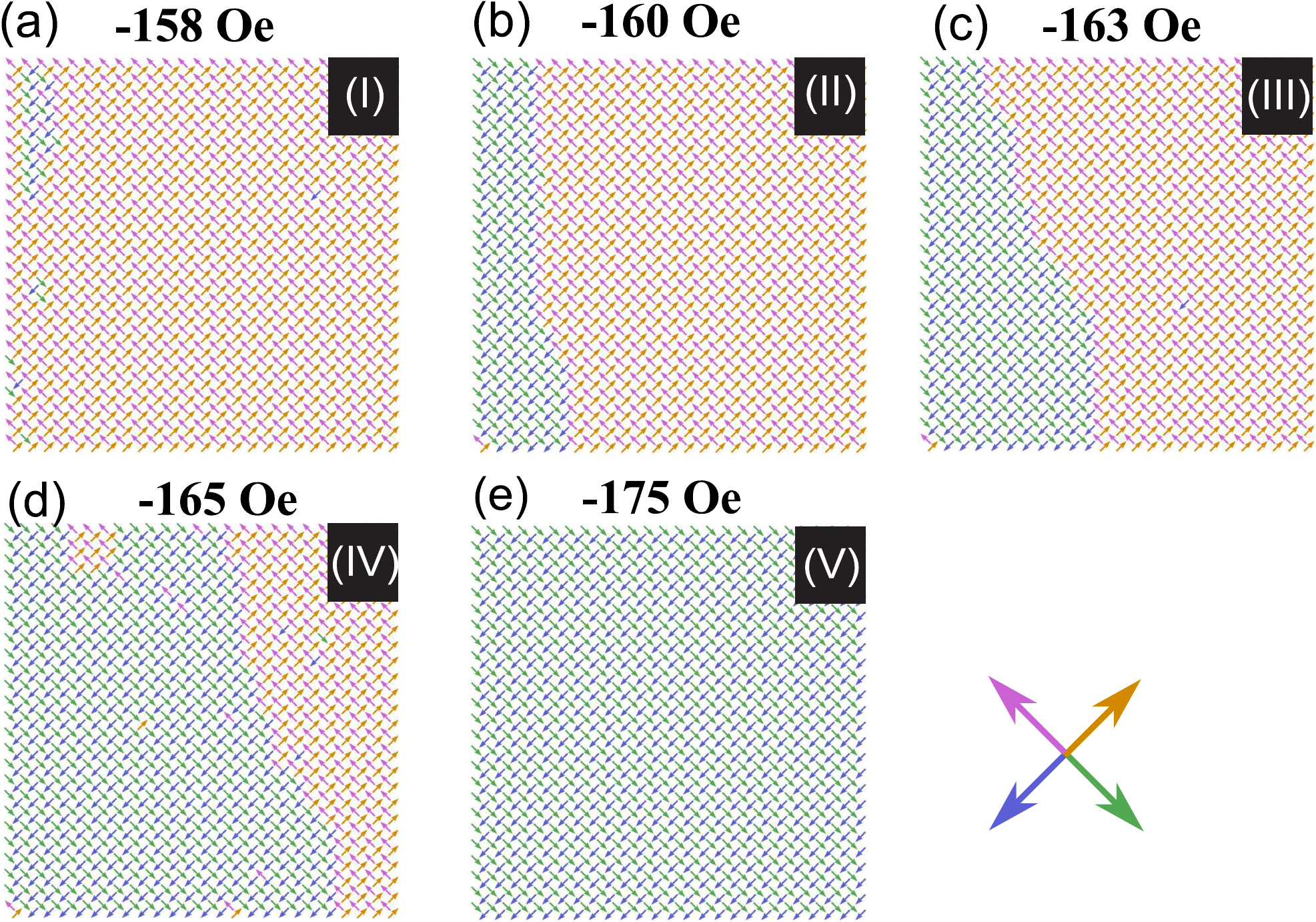}
	\caption{\textbf{Visulisation of the field-induced magnetisation maps of individual nanomagnets at $-6^\circ$ magnetic field}. \textbf{(a)-(e)} Evolution of macrospins flipping as a function of magnetic field with the strength of \textbf{(a)} -158~Oe, \textbf{(b)} -160~Oe, \textbf{(c)} -163~Oe, \textbf{(d)} -165~Oe and \textbf{(e)} 175~Oe. The colour code on the bottom right indicates the magnetisation direction of each coloured arrow. }
	\label{MagIslandPattern}
\end{figure}

In Figs.~\ref{DomainPattern} and \ref{DomainPattern30} of the main text we only looked at the behaviour of the unit net moment.
In Fig.~\ref{MagIslandPattern} we give additional images showing the reversal of the individual islands extracted from the LTEM images.
In particular, we highlight that the 2-D nature of the reversal is a consequence of the simultaneous switching of the sublattices. 

\newpage

\section*{Reversal Videos} \label{DWform}

\textbf{Video S1.}  Video showing how the magnetisation of each nanomagnet (left panel) and net moments of the pinwheel units (right panel) reverse on the ‘down’ branch in the vicinity of coercivity at $-6^\circ$ field with respect to the array. 

\noindent \textbf{Video S2.} Video showing how the magnetisation of each nanomagnet (left panel) and net moments of the pinwheel units (right panel) reverse on the ‘down’ branch in the vicinity of coercivity at $30^\circ$ field with respect to the array.

\newpage

\section*{Domain Wall Population} \label{wallpop}
\begin{figure}[ht]
\centering
\includegraphics[width=0.9\linewidth]{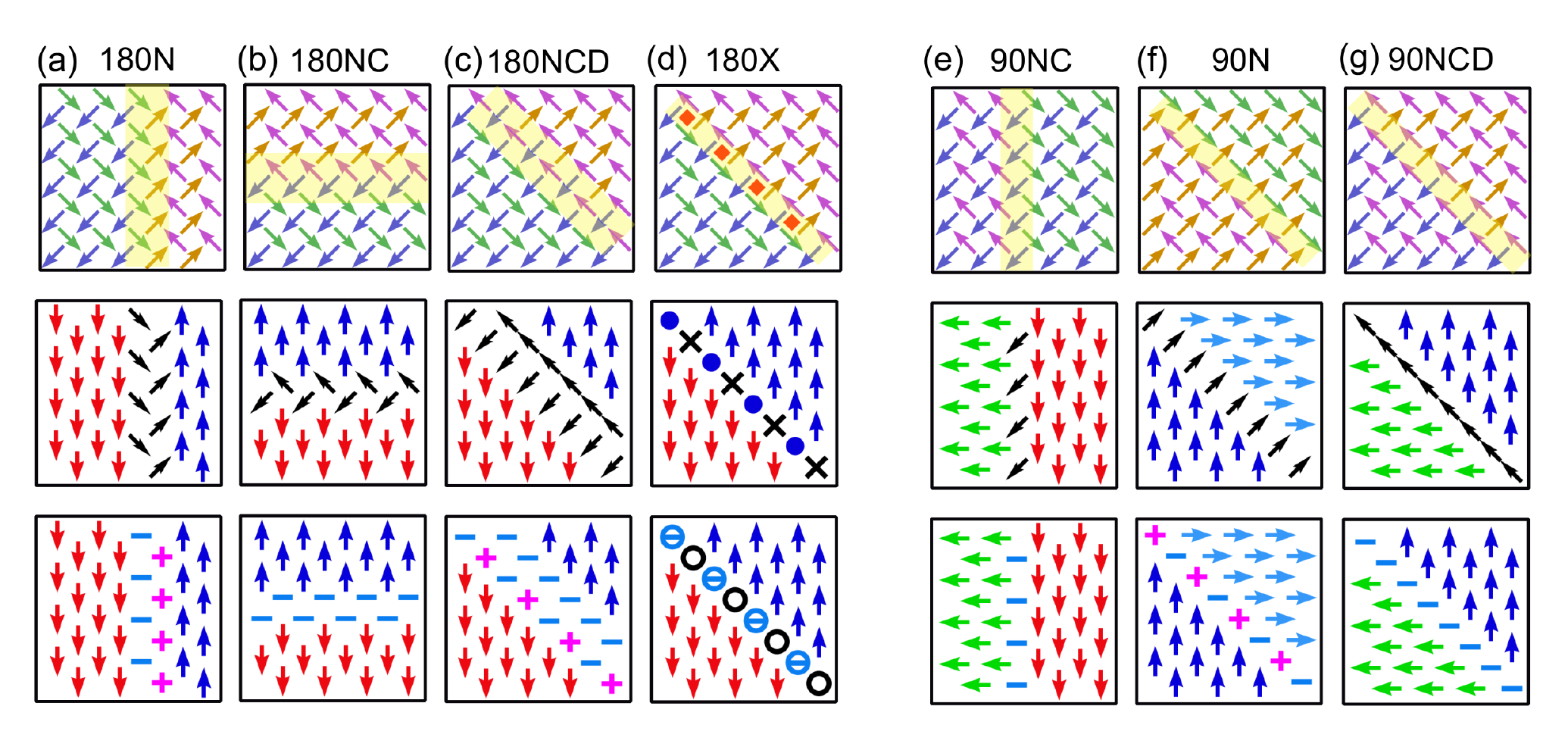}
\caption{\textbf{Magnetisation and charge-ordering topologies of other possible mesoscopic DWs}. \textbf{(a)-(d)} Schematic examples of possible $180^\circ$ DW configurations in the pinwheel ASI containing four categories: \enquote*{180N} \textbf{(a)}, \enquote*{180NC} \textbf{(b)}, \enquote*{180NCD} \textbf{(c)} and \enquote*{180X} \textbf{(d)}. \textbf{(e)-(g)} Schematics of possible $90^\circ$ DW configurations consisting of three types: \enquote*{90NC} \textbf{(e)}, \enquote*{90N} \textbf{(f)} and \enquote*{90NCD} \textbf{(g)}. }
\label{DW2}
\end{figure}

We note that the domain wall definitions shown in Fig.~\ref{DWs} of the main text are not the only possibilities.
Another set of possible domain wall configurations are shown in Fig.~\ref{DW2}  for the asymmetric array.

\begin{figure}[ht]
	\centering
	\includegraphics[width=0.9\linewidth]{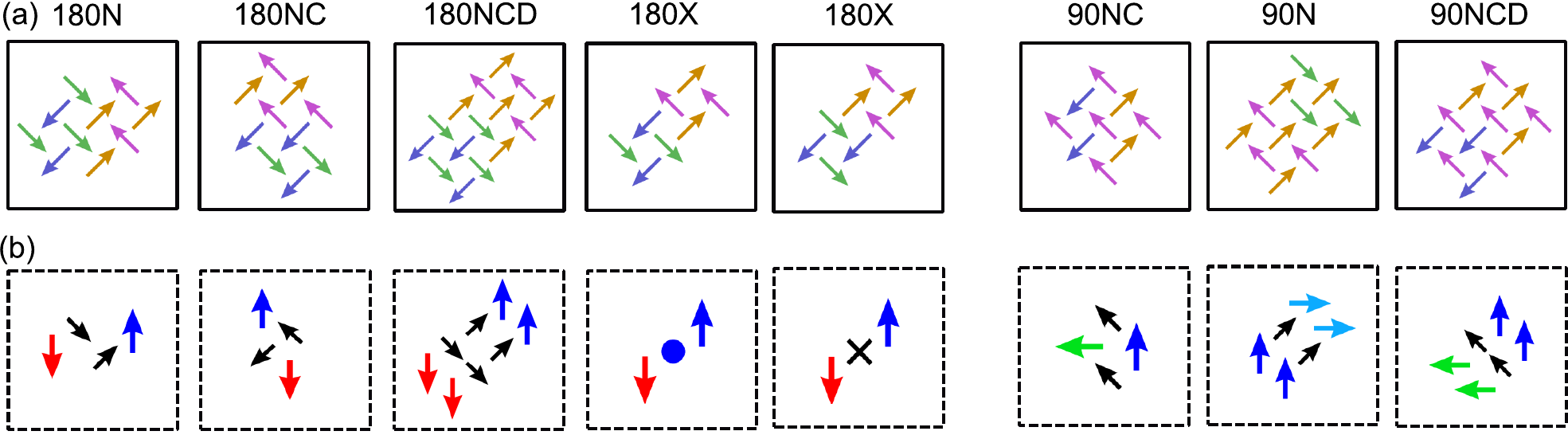}
	\caption{\textbf{Domain wall motifs in pinwheel ASI}. \textbf{(a)} Example minimum domain wall motifs and \textbf{(b)} unit magnetisation topologies for the different magnetisation configurations. Other possible DW motifs exist but, for brevity, are not shown here.}
	\label{DWs_unit}
\end{figure}

In Fig.~\ref{DWs_unit}, we show the domail wall motifs used to calculate the domail wall population statistics discussed in the main text.
To uniquely define each wall, it is necessary to extend the motifs of some domain walls along the `length' of the wall.
For example, the 180NCD, 90N, 90NC and 90NCD motifs are twice the length of the others and thus are weighted twice the amount of other motifs when counting the DW populations.

\begin{figure}[ht]
	\centering
	\includegraphics[width=0.8\linewidth]{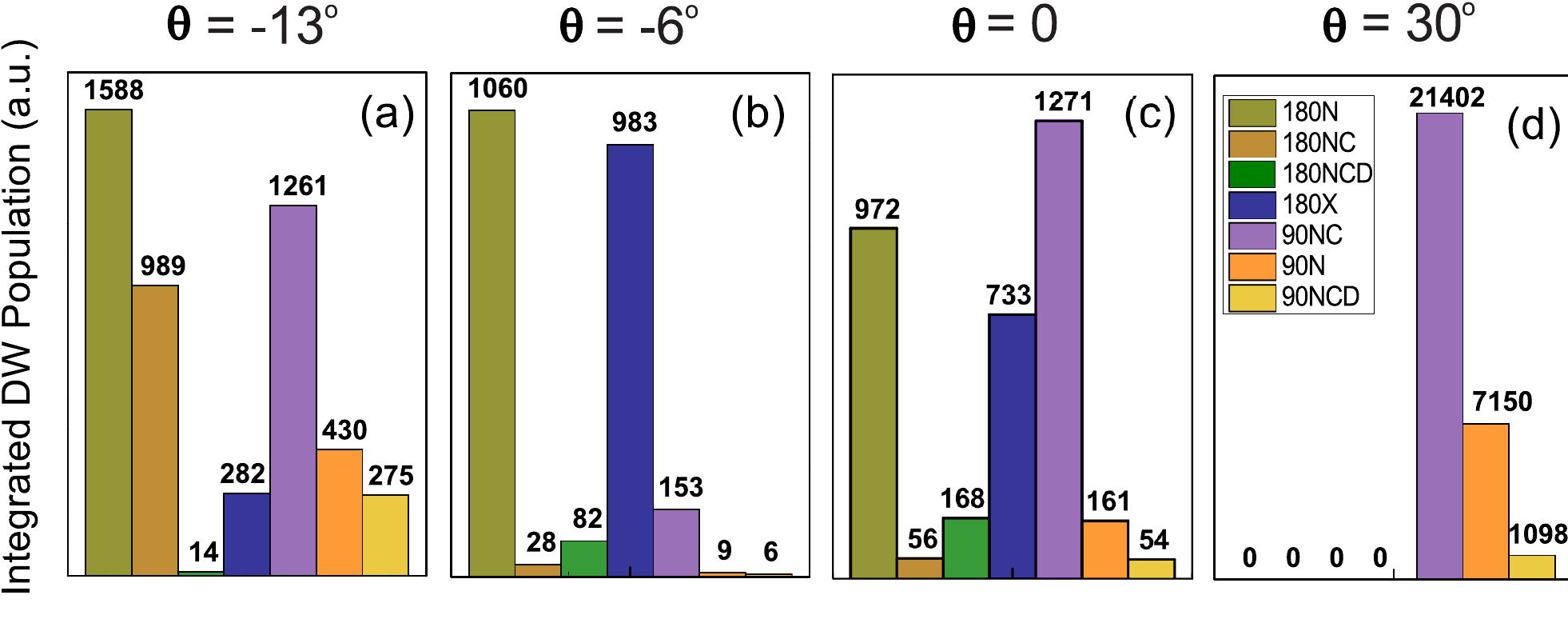}
	\caption{\textbf{Complete domain wall populations during an M-H loop.} \textbf{(a)-(b)} Domain wall population histograms for the seven domain wall classes discussed in the main text, at the indicated applied field angles, matching those used in the polar magnetisation loops. The number above each column indicates the integrated population of domain wall motifs over a complete M-H loop.}
	\label{DW_Pop_absolute}
\end{figure}

In Fig.~\ref{DW_Pop} of the main text we have shown the angular dependence of the integrated domain wall population around a hysteresis loop.
This can be decomposed into the populations of each of the seven domain wall types at each applied field angle.
In Fig.~\ref{DW_Pop_absolute} we show examples of the individual integrated domain wall populations at the relevant applied field angles.

\end{document}